\documentclass[
 aip,
 amsmath,amssymb,
 reprint,
]{revtex4-2}

\usepackage{graphicx} 
\usepackage{dcolumn} 
\usepackage{bm} 
\usepackage[utf8]{inputenc}
\usepackage[T1]{fontenc}
\usepackage{mathptmx}
\usepackage{etoolbox}
\usepackage{textgreek}
\usepackage{xcolor}
\usepackage{upgreek}

\makeatletter
\def\@email#1#2{
 \endgroup
 \patchcmd{\titleblock@produce}
  {\frontmatter@RRAPformat}
  {\frontmatter@RRAPformat{\produce@RRAP{*#1\href{mailto:#2}{#2}}}\frontmatter@RRAPformat}
  {}{}
}
\makeatother

\begin{document}

\preprint{AIP/123-QED}

\title[Gamma-Flash by Laser-Matter Interaction]{Gamma-Flash Generation in Multi-Petawatt Laser-Matter Interactions}

\author{P. Hadjisolomou}
\email[Electronic mail: ]{Prokopis.Hadjisolomou@eli-beams.eu} 
\affiliation{ELI Beamlines Facility, Extreme Light Infrastructure ERIC, Za Radnicí 835, 25241 Dolní Břežany, Czech Republic}

\author{T. M. Jeong}
\affiliation{ELI Beamlines Facility, Extreme Light Infrastructure ERIC, Za Radnicí 835, 25241 Dolní Břežany, Czech Republic}

\author{D. Kolenaty}
\affiliation{Department of Physics and NTIS - European Centre of Excellence, University of West Bohemia, Univerzitní 8, 306 14 Plzeň, Czech Republic}

\author{A. J. Macleod}
\affiliation{ELI Beamlines Facility, Extreme Light Infrastructure ERIC, Za Radnicí 835, 25241 Dolní Břežany, Czech Republic}

\author{V. Olšovcová}
\affiliation{ELI Beamlines Facility, Extreme Light Infrastructure ERIC, Za Radnicí 835, 25241 Dolní Břežany, Czech Republic}

\author{R. Versaci}
\affiliation{ELI Beamlines Facility, Extreme Light Infrastructure ERIC, Za Radnicí 835, 25241 Dolní Břežany, Czech Republic}

\author{C. P. Ridgers}
\affiliation{York Plasma Institute, Department of Physics, University of York, Heslington, York, North Yorkshire YO10 5DD, UK}

\author{S. V. Bulanov}
\altaffiliation[Also at ]{National Institutes for Quantum and Radiological Science and Technology (QST), Kansai Photon Science Institute, 8-1-7 Umemidai, Kizugawa, Kyoto 619-0215, Japan}
\affiliation{ELI Beamlines Facility, Extreme Light Infrastructure ERIC, Za Radnicí 835, 25241 Dolní Břežany, Czech Republic}

\date{\today}

\begin{abstract}
The progressive development of high power lasers over the last several decades, enables the study of \textgamma-photon generation when an intense laser beam interacts with matter, mainly via inverse Compton scattering at the high intensity limit. \textgamma-ray flashes are a phenomenon of broad interest, drawing attention of researchers working in topics ranging from cosmological scales to elementary particle scales. Over the last few years, a plethora of studies predict extremely high laser energy to \textgamma-photon energy conversion using various target and/or laser field configurations. The aim of the present manuscript is to discuss several recently proposed \textgamma-ray flash generation schemes, as a guide for upcoming \textgamma-photon related experiments and for further evolution of the presently available theoretical schemes.
\end{abstract}

\maketitle


\section{Introduction}

Development of multi-PW lasers \citep{2022_RadierC, 2021_YoonJ, 2016_PapadopoulosDN, 2019_DansonC} paves the path towards the possibility of the experimental investigation of the radiation dominated regime where radiation reaction and quantum effects play a significant role in particle dynamics. Today, laser power of $10 \kern0.2em \mathrm{PW}$ has been reported \citep{2022_RadierC}, whilst laser intensity exceeding $10^{23} \kern0.2em \mathrm{W cm^{-2}}$ has been reached \citep{2021_YoonJ}.

Intense laser-solid interactions result in several distinct populations of energetic \textgamma-photons, leptons (electrons, positrons) and hadrons (protons, ions). Under the effect of strong fields electrons and positrons quickly become relativistic due to their relatively low mass. As electric field is not a Lorentz invariant, the electrons in their boosted frame can approach or even surpass the Scwinger field, $E_S$ \citep{1931_SauterF, 1936_HeisenbergW, 1951_SchwingerJ}; $E_S$ is the limit of strong fields at which spontaneous electron-positron ($e^-e^+$) pair generation can occur due to vacuum breakdown. Therefore, QED effects can be observed even with currently available lasers.

Although particle-in-cell (PIC) codes were developed many decades ago, QED PIC routines have been developed for major PIC codes only in the last decade. These routines allow extensive computational investigation of the collective behavior governing the QED effects in a multi-species plasma which is inevitably generated when the strong laser field interacts with matter. Two main processes dominate at ultra-high intensities and are implemented in the QED PIC codes, being the emission of a \textgamma-photon through the multiphoton (inverse) Compton scattering \citep{1964_NikishovAI, 1964_BrownLS, 1964_GoldmanII, 1985_RitusVI} and the generation of an $e^-e^+$ pair through the multiphoton Breit-Wheeler \citep{1934_BreitG, 1962_ReissHR, 1964_NikishovAI, 1966_YakovlevVP} process.

Here, we focus on Compton \textgamma-photon emission. Bright \textgamma-photon sources find applications in a plethora of topics, including but not limited to study of fundamental physics \citep{2016_HommaK}, positron generation \citep{2000_GahnC, 2015_SarriG, 2022_KolenatyD}, neutron sources \citep{2014_PomerantzI} astrophysical studies \citep{2015_BulanovSV, 2015_SarriG}, photonuclear fission \citep{2000_CowanTE, 2003_SchwoererH}, radiotherapy \citep{1997_weeksKJ}, fine measurement of atomic nuclei \citep{2014_TarbertCM} and shock-wave studies \citep{2008_RavasioA, 2017_AntonelliL}.


\section{QED Processes in Strong Fields}


The interaction of an electromagnetic field with charged particles is conveniently described by the normalized dimensionless electromagnetic wave amplitude \citep{2006_MourouGA},
\begin{equation}
a_0 = \frac{e E_0}{m_e \omega_l c} .
\label{eq:B4}
\end{equation}
Here, $e$ is the elementary charge, $E_0$ is the electric field amplitude, $m_e$ is the electron rest mass, $c$ is the speed of light in vacuum and $\omega_l=2 \pi c/\lambda_l$ is the laser frequency (wavelength $\lambda_l$). When $a_0 = 1$ the work done by electric field $E_0$ acting on a particle of charge $e$ over a distance of $\lambda / (2 \pi)$, equals $m_e c^2$. An electron oscillating in the laser field is therefore considered as relativistic when $a_0 \geqslant 1$, since then the electron obtains velocities similar to $c$ during its oscillatory motion. The laser intensity is connected to $a_0$ by
\begin{equation}
I_L = \frac{\varepsilon_0 \omega_l^2 c^3 a_0^2}{2 e^2} \approx 1.37 \times 10^{18} \left( \frac{1 \mathrm{\upmu m}}{\lambda_l} \right)^2 a_0^2 \kern0.2em \mathrm{W cm^{-2}} ,
\label{eq:B4a}
\end{equation}
where $\varepsilon_0$ is the vacuum permittivity.

A dimensionless parameter describing the radiation losses is
\begin{equation}
\varepsilon_{rad} = \frac{4 \pi r_e}{3 \lambda_l} ,
\label{eq:B31}
\end{equation}
where $r_e = e^2/(4 \pi \varepsilon_0 m_e c^2)$ is the classical electron radius. Radiation reaction refers to the recoil force experienced by an accelerated particle emitting radiation. Radiation effects dominate \citep{2004_BulanovSV, 2011b_BulanovSV} for
\begin{equation}
a_0 \varepsilon_{rad}^{1/3} > 1 .
\label{eq:B31a}
\end{equation}

\subsection{Intensity Levels}


In this subsection let us assume a $1 \mathrm{\mu m}$ wavelength laser. By requiring energy balance \citep{2004_BulanovSV} between the power emitted by the electron, $P_\gamma \approx \varepsilon_{rad} m_e c^2 \omega_l \gamma_e^4$, and the electron energy acquisition rate, $P_{acq} = m_e c^2 \omega_l \gamma_e$, then $a_0 = \varepsilon_{rad}^{-1/3}$ corresponds to an intensity of
\begin{equation}
I_R = \left( \frac{3}{2} \right)^{2/3} \frac{m_e^{8/3} c^5 \omega_l^{4/3}}{2 \pi (e/\sqrt{4 \pi \varepsilon_0})^{10/3}} ,
\label{eq:B32}
\end{equation}
which gives an intensity of $2.64 \kern0.1em {\times} \kern0.1em 10^{23} \kern0.2em \mathrm{W cm^{-2}}$.


When the energy of the emitted \textgamma-photon energy is comparable to the electron energy, then $\hbar \omega_\gamma \approx \gamma_e m_e c^2$ and QED effects become important. A rotating electron in a circularly polarized laser field emits photons of energy $\hbar \omega_\gamma \approx \hbar \omega_l \gamma_e^3$ \citep{2015_BulanovSV}. Therefore, $\gamma_e = \sqrt{m_e c^2 / (\hbar \omega_l)} \approx 642$. For laser intensities greater than $10^{23} \kern0.2em \mathrm{W cm^{-2}}$, the electron energy scales as $m_e c^2 (a_0 / \varepsilon_{rad})^{1/4}$. As a result, \citep{2004_BulanovSV, 2011a_BulanovSV} in the QED regime
\begin{equation}
 a_0 = \frac{2 e^2/(4 \pi \varepsilon_0) m_e c}{3 \hbar \omega_l} = \left( \frac{2}{3} \alpha \right)^2 \varepsilon_{rad}^{-1} ,
\label{eq:B2_6}
\end{equation}
($\alpha = e^2/(4 \pi \varepsilon_0 \hbar c)$ is the fine structure constant) which gives $a_0 \approx 2005$, corresponding to an intensity of $1.1 \kern0.1em {\times} \kern0.1em 10^{25} \kern0.2em \mathrm{W cm^{-2}}$.


The Schwinger field \citep{1931_SauterF, 1936_HeisenbergW, 1951_SchwingerJ, 2015_BulanovSV} is defined as the field required to accelerate an electron to its rest mass energy, over a Compton wavelength, $\lambda_C = 2 \pi \hbar / m_e c$. The Schwinger field is given by
\begin{equation}
 E_S = \frac{m_e^2 c^3}{e \hbar}
       \approx 1.32 \kern0.1em {\times} \kern0.1em 10^{18} \kern0.2em \mathrm{V m^{-1}} ,
\label{eq:B35}
\end{equation}
corresponding to an intensity of
\begin{equation}
I_S = \frac{m_e^4 c^7}{2 \pi (e/\sqrt{4 \pi \varepsilon_0})^2 \hbar^2}
     \approx 2.33 \kern0.1em {\times} \kern0.1em 10^{29} \kern0.2em \mathrm{W cm^{-2}} .
\label{eq:B36}
\end{equation}
At this limit, virtual $e^-e^+$ pairs in vacuum can be separated, producing real $e^-e^+$ in a process known as Sauter–Schwinger \citep{1931_SauterF, 1936_HeisenbergW, 1951_SchwingerJ} pair production.



\subsection{The $\chi_e$ and $\chi_\gamma$ Parameters} \label{chieg}

The relativistic gauge invariant parameter for electron is $\chi_e$,
\begin{equation}
\chi_e =  \sqrt{ \left(\gamma_e \frac{{\bm{E}}}{E_S} + \frac{{\bm{p}}}{m_e} \kern0.1em {\kern0.1em {\times} \kern0.1em } \kern0.1em  \frac{{\bm{B}}}{E_S} \right)^2 - \left(\frac{{\bm{p}}}{m_e c} \cdot \frac{{\bm{E}}}{E_S} \right)^2 } ,
\label{eq:xi-e}
\end{equation}
and it characterizes the importance of quantum effects in \textgamma-photon emission. If $\chi_e \gg 1$ then the Compton scattering \textgamma-photon emission is a multiphoton process,
\begin{equation}
 e^\pm + N \omega_l \rightarrow e^\pm + \omega_\gamma .
\label{eq:2011B68}
\end{equation}
In the electron rest frame, the scattered \textgamma-photon frequency is related to that of the laser \citep{2009_EhlotzkyF, 2011a_BulanovSV} as
\begin{equation}
\omega_\gamma = \frac{N \omega_l}{1+[N \hbar \omega_l/(m_e c^2)+a_0^2/4][1-\cos(\theta)]} ,
\label{eq:2011B69}
\end{equation}
with $\theta$ being the angle between the laser and \textgamma-photon propagation directions.


The dimensionless Lorentz invariant parameter for \textgamma-photon is $\chi_\gamma$,
\begin{equation}
\chi_\gamma =  \frac{\hbar \omega_l}{m_e c^2} \sqrt{ \left( \frac{{\bm{E}}}{E_S} + c {\bm{\hat{p}}} \kern0.1em {\kern0.1em {\times} \kern0.1em } \kern0.1em  \frac{{\bm{B}}}{E_S} \right)^2 - \left({\bm{\hat{p}}} \cdot \frac{\bm{E}}{E_S} \right)^2 } .
\label{eq:xi-g}
\end{equation}
and for larger $\chi_\gamma$ values the probability of the multiphoton Breit-Wheele $e^-e^+$ pair generation is increased,
\begin{equation}
\gamma + N_l \omega_l \rightarrow e^- + e^+ .
\label{eq:B93}
\end{equation}
During that process, a \textgamma-photon interacts with $N_l$ laser photons to produce an $e^-e^+$ pair \citep{2011a_BulanovSV}, where $N_l = a_0 \chi_\gamma^{-1}$.


\section{Implementation of Monte-Carlo Method for QED PIC Codes}


\textgamma-photon emission and $e^-e^+$ pair generation can be described by the Monte-Carlo method \citep{2009_EhlotzkyF, 2010_SokolovIV}, thus capturing the statistical nature of these processes.  While there are several ways to implement such a Monte-Carlo algorithm, a computationally efficient method makes use of the cumulative probability of emission.  The EPOCH QED module \citep{2012_RidgersCP, 2013_RidgersCP, 2014_RidgersCP} is an example of this method. For a given particle, a random positive number less than or equal to unity is set to the cumulative probability at which emission will occur, $P (t) = 1 - \exp(- \tau_{em})$.  Here $\tau_{em}$ is the optical depth at which emission occurs and the previous equation is then inverted to yield the value of $\tau_{em}$ at which the particle will emit.  For each particle the optical depth is updated according to $\tau(t) = \int \lambda dt$, where here $\lambda$ is the rate of the emission process. Emission occurs when $\tau=\tau_{em}$.

  The required rates of \textgamma-photon emission and $e^-e^+$ pair generation are given by assuming that the electric and magnetic fields are locally constant and crossed; for discussion on the validity of this approach see e.g. \citep{1985_RitusVI, 1989_BaierVN, 2016_DimuV, 2018_DiPiazzaA, 2019_IldertonA, 2020_HeinzlT}. If the momentum transferred to the background field is ignored, then the emitted photon momentum is balanced by the recoil of the emitting electron (or positron). When a \textgamma-photon generates an $e^-e^+$ pair its energy is shared between the electron and positron.

  In a PIC code \citep{2002_FonsecaRA, 2015_ArberTD, 2018_DerouillatJ} many real particles are represented by a single macroparticle in order that the computational cost of simulating the particles is not prohibitive. The momentum and position of each macroparticle is obtained by integration of the relativistic Lorentz equation. Subsequently, the charge and current values are obtained by interpolating the macroparticle positions and velocities onto a computational grid. These current and charge densities are then used to update the electric and magnetic field values in each cell via a discretization of Maxwell's equations. These updated fields are then used to update the momentum and position of the macroparticles in the next time step and so on in a computational loop.

  The Monte-Carlo approach to simulating \textgamma-photon emission and $e^-e^+$ pair generation lends itself naturally to inclusion in a PIC code. In the resulting QED-PIC approach the electromagnetic fields are  split into low (e.g. laser field) and high frequency (\textgamma-ray) components \citep{2014_RidgersCP}. The low frequency component is determined by solving Maxwell's equations in the usual way.  The high frequency \textgamma-ray component is discretized as macroparticles which propagate ballistically.  The emission of these \textgamma-ray macroparticles, by electrons and positrons (nonlinear inverse Compton scattering), as well as their subsequent generation of electron positron pairs (nonlinear Breit-Wheeler process), follows the Monte-Carlo method outlined above \citep{2010_DuclousR, 2012_RidgersCP}. More recently, this method has also been used to describe Bremsstrahlung \citep{2019_ReesA,2019_Martinez,2021_MorrisS} and Bethe-Heitler pair production \cite{2019_Martinez}.

  In addition to the Monte-Carlo approach, a semi-classical model for \textgamma-ray emission by nonlinear inverse Compton scattering has been included in some PIC codes.  Here the electrons and positrons radiate continuously but the radiated intensity $I$ is reduced by the Gaunt factor $G(\chi_e)=I/I_{cl}$ (where $I_{cl}$ is the intensity radiated in a classical model) \citep{2008_BellAR, 2009_KirkJG, 2010_DuclousR, 2014_RidgersCP}. A semi-classical model accurately captures the evolution of averaged quantities such as athe average energy loss, but it is less accurate than the Monte-Carlo approach in describing quantities influenced by the stochasticity \cite{2014_RidgersCP, 2018_NielF} of the QED processes. It can be useful, however, as a comparison to the Monte-Carlo model to reveal the effects of quantum stochasticity, and in the development of theoretical models, where the inclusion of deterministic radiation is far more straightforward \cite{2015_ZhangP, 2018_DelSorboD}.





\section{Particle-in-Cell $\bold{\gamma}$-photons}

At $a_0 \approx 50$, radiation damping effects become important \citep{1996_HartemannFV, 2002_ZhidkovA}. The following discussion is based on QED PIC simulations of femtosecond-class lasers focused to a micrometer scale focal spot interacting with matter, except where it is otherwise stated. The simulations are either one-dimensional or two-dimensional, except where 3D simulations are specified. These results, based on the interaction of ultra-intense lasers with ever more innovative targets, illustrate the continuously evolving effort to enhance the conversion efficiency of the laser energy into \textgamma-photon energy, $\kappa_\gamma$.


\subsection{Foil Targets}

A $10^{22} \kern0.2em \mathrm{W cm^{-2}}$ intensity laser emits $0.2 \kern0.2em \%$ of its energy as photons \citep{2002_ZhidkovA} when interacting with a $3 \mathrm{\mu m}$ thick copper slab proceeded by an exponential preplasma. The preplasma gradient strongly affects the laser absorption efficiency. Firstly, electrons are accelerated by the ${\bf{E}} \kern0.1em {\times} \kern0.1em {\bf{B}}$ force. The reflected laser field results in the second electron population. The third population is due to compensation of the ${\bf{E}} \kern0.1em {\times} \kern0.1em {\bf{B}}$ force by the longitudinal plasma field, where return currents (electrons) further increase the emitted radiation. When increasing the laser intensity by an order of magnitude, $\kappa_\gamma$ increases to $20 \kern0.2em \%$, fitted by $\kappa_\gamma \propto I^2$ for the intensities used.

A similar setup \citep{2012_NakamuraT} uses an an overcritical $10 \kern0.2em \mathrm{\mu m}$ thick foil with an added preplasma, interacting with a laser of $a_0 = 150$. The resulting \textgamma-ray flash duration is comparable to that of the laser pulse. Existence of an optimal preplasma scale-length is found, with a long preplasma being less prone to laser back-reflection. A  $\kappa_\gamma \approx 32 \kern0.2em \%$ is estimated, attributed to self-focusing of the laser which leads to enhancement of the field and consequent high laser depletion. Moreover, the emitted radiation occurs in two symmetric lobes.

When a $4 \kern0.1em {\times} \kern0.1em 10^{23} \kern0.2em \mathrm{W cm^{-2}}$ intensity laser interacts with a $1 \kern0.2em \mathrm{\mu m}$ thick aluminium foil \citep{2012_RidgersCP, 2013_RidgersCP}, the laser pressure bores a cavity into the foil, where both \textgamma-photons and $e^-e^+$ pairs are created. A preplasma is found to enhance the \textgamma-photon yield. The simulations result in $\kappa_\gamma \approx 10 \kern0.2em \%$ for a $12.5 \kern0.2em \mathrm{PW}$ laser and $\kappa_\gamma \approx 40 \kern0.2em \%$ for a $320 \kern0.2em \mathrm{PW}$ laser.
Optimization of \textgamma-photon yield relies on a preplasma multiparametric study \citep{2018_LezhninKV}, where the variables include the laser power, pulse duration, focal spot, target thickness and preplasma exponent factors. The optimal \textgamma-flash generation occurs when the laser undergoes self-focusing in the preplasma. When the laser propagates within low density, the generated \textgamma-photons are emitted backwards from electrons counter-propagating relative to the laser. This tendency reverses at denser preplasma regions. 3D simulations report $\kappa_\gamma \approx 20 \kern0.2em \%$ for the $10 \kern0.2em \mathrm{PW}$ case.

The importance of the proper electron density and foil thickness choice on optimizing \textgamma-photon generation \citep{2020_MartinezB} is illustrated by the interaction of a $10^{22} \kern0.2em \mathrm{W cm^{-2}}$ intensity laser with a foil having an electron density of either $1.87 \kern0.1em {\times} \kern0.1em 10^{22} \kern0.2em \mathrm{cm^{-3}}$ (relativistically transparent, undercritical) or $1.1 \kern0.1em {\times} \kern0.1em 10^{23} \kern0.2em \mathrm{cm^{-3}}$ (opaque, overcritical). For the undercritical target case, \textgamma-photon production is associated with energetic electrons counter-propagating with the laser. The spatial distribution of the radiated power reveals periodic modulations attributed to relativistic Doppler effects, resulting in \textgamma-ray bursts towards the target front. The \textgamma-photon yield diminishes for foils thinner than $1 \kern0.2em \mathrm{\mu m}$ due to rapid target expansion. For overcritical targets, \textgamma-photon generation occurs in regions defined by the expanding preplasma and/or the skin-depth layer. For a relatively thick foil, \textgamma-photon emission occurs mainly at a broad angle with respect to the laser axis , in the forward direction, as a result of back-directed electrons to the target. For foils thinner than $1 \kern0.2em \mathrm{\mu m}$, a maximal emission angle occurs at both the forward and backward directions, attributed to recirculating electrons.

For a comparable (to \citep{2020_MartinezB}) simulation setup \citep{2020_VyskocilJ}, at a fixed density of  $3.18 \kern0.1em {\times} \kern0.1em 10^{23} \kern0.2em \mathrm{cm^{-3}}$, but with various intensity values covering the range of $3 \kern0.1em {\times} \kern0.1em 10^{21} \kern0.2em \mathrm{W cm^{-2}} \text{\textendash} 10^{23} \kern0.2em \mathrm{W cm^{-2}}$, the double-lobe pattern is reaffirmed. The lobes are maximized at $\sim 30^o$, with the exact value depending on the target thickness \citep{2020_VyskocilJ}. For targets thicker than $2 \kern0.2em \mathrm{\mu m}$ the \textgamma-photon angular distribution saturates. A theoretical model \citep{2020_VyskocilJ} estimates the \textgamma-photon emission angle by assuming that the incoming and reflected parts of the laser pulse form a planar standing wave. Within the intensity range examined, the \textgamma-photon spectrum temperature increases linearly to the laser intensity.

The double-lobe \textgamma-photon distribution is also observed \citep{2014_JiLL} on simulations of a of $5.4 \kern0.1em {\times} \kern0.1em 10^{23} \kern0.2em \mathrm{W cm^{-2}}$ intensity laser interacting with a relativistically undercritical target. The laser penetrates the target forming a cavity. A periodic electron distribution is formed on the laser axis, dragging protons and forming quasineutral bunches. The radiation reaction force reaches values comparable to the ponderomotive force that classically expels electrons from the cavity region, and electrons are trapped in the laser field region. The sequential recoil of those trapped electrons results in \textgamma-photons emitted at two directions ($15^o-30^o$ and $330^o-345^o$), with $\kappa_\gamma \approx 35 \kern0.2em \%$.

The \textgamma-photon emission angle has also been investigated through the transition \citep{2018_DuffMJ, 2019_PopruzhenkoSV} from the Light-Sail (thin targets) to the Hole-Boring (thick targets) regime, where a $2 \kern0.1em {\times} \kern0.1em 10^{23} \kern0.2em \mathrm{W cm^{-2}}$ intensity laser interacts with an aluminum target, varying in thickness from $50 \kern0.2em \mathrm{nm}$ to $400 \kern0.2em \mathrm{nm}$. The Hole-Boring regime allows strong electron return currents resulting in \textgamma-photon emission. Due to strong charge separation fields, the longitudinal electron momentum acquires large values, decreasing the \textgamma-photon emission angle.

\subsection{Structured Targets} \label{ST}

In the relativistically transparent regime, at non-optimal plasma densities the laser deviates from its propagation axis and the emitted \textgamma-photons are non-directional. To overcome this issue \citep{2016_StarkDJ}, an undercritical channel is added to an overcritical target that interacts with a $5 \kern0.1em {\times} \kern0.1em 10^{22} \kern0.2em \mathrm{W cm^{-2}}$ intensity laser. A slowly varying magnetic field appears within the channel. Due to self-generation of the magnetic field, electrons moving within the channel feel an enhanced accelerating force, increasing the \textgamma-photon yield. The laser-target interaction under this scheme demonstrates $\kappa_\gamma \approx 15 \kern0.2em \%$. The emission occurs near the channel walls, at the electron trajectory turning points. Therefore, \textgamma-photon emission occurs at \textgamma-ray bursts of period same as that of the laser. 3D simulations verified the two-lobed \textgamma-photon emission pattern, although $\kappa_\gamma$ dropped to $3.5 \kern0.2em \%$.

In the channel targets, by increasing the laser power the deviation of the lobes from the laser axis decreases \citep{2020_WangT}. For a $4 \kern0.2em \mathrm{PW}$ laser $\kappa_\gamma \approx 1.5 \kern0.2em \%$. The number of \textgamma-photons is proportional to $P^2$ for $P < 4 \kern0.2em \mathrm{PW}$, and proportional to $P$ for $P > 4 \kern0.2em \mathrm{PW}$. A model finding optimal conditions for \textgamma-photon emission by channel targets has been constructed \citep{2021_RinderknechtHG}, where the number of \textgamma-photons scales proportionally to the laser energy, and $\kappa_\gamma \propto a_0^3$. If the cylindrical target is combined with Poincar\'e beams \citep{2022_YounisD}, emission of collimated \textgamma-photons is expected.

In a diametrically opposite target design \citep{2018_WangWM}, irradiation of a 3D long thin wire by a $2.5 \kern0.1em {\kern0.1em {\times} \kern0.1em } \kern0.1em  10^{22} \mathrm{W cm^{-2}}$ intensity laser is proposed for efficient \textgamma-photon generation. The findings suggest $\kappa_\gamma \approx 10 \%$. The rectangular aluminum wire has a $0.6 \mathrm{\mu m}$ edge. The laser-wire interaction initiated before the laser reached the focal spot, to gradually accelerate electrons along the wire direction with their divergence reduced, resulting in the existence of a collimated electron population at the time when the pulse reaches its highest intensity on the focal spot. The resulting \textgamma-photons are highly collimated, with a maximum energy of $500 \mathrm{MeV}$.

A hybrid scheme \citep{2020_ZhuXL} for \textgamma-photon emission, as a first stage, employs a channel filled with a low plasma density.  The density has a transversely paraboloidal profile and so propagation of the laser in the channel results in a collimated high energy, high density electron beam. The second stage employs a higher density plasma region, attached to the low density plasma channel. In the vicinity of the electron beam, strong electrostatic fields are formed, corresponding to $\chi_e \approx 0.1$. The high $\chi_e$ value of the directional electron beam results in a collimated \textgamma-photon beam, where for a $10 \mathrm{PW}$ laser a $\kappa_\gamma \approx 16 \%$ is reported.

A 3D scheme based on reflection of a circularly polarized laser \citep{2021_ZhangH} of $1.4 \kern0.1em {\times} \kern0.1em 10^{22} \kern0.2em \mathrm{W cm^{-2}}$ intensity employs a cylindrical micro-channel target attached to a square foil consisting of eight triangular segments (height difference of $0.25 \pi \lambda_l$) joined in the center of the square. The triangular segments result in a phase difference of the reflected pulse and transformation of the beam to a Laguerre-Gaussian beam. Prior to reflection, electrons are dragged out from the micro-channel walls and accelerated by the longitudinal field formed in the channel, where the micro-channel acts as an optical waveguide. Interaction of the accelerated electrons with the reflected (to the square foil) pulse results in a collimated \textgamma-photon beam with a $\kappa_\gamma \approx 1.2 \%$.

Micro-fabrication of a conical structure attached to a foil target is proposed \citep{2022_ChintalwadS}, interacting with a $2 \kern0.1em {\times} \kern0.1em  10^{22} \kern0.2em \mathrm{W cm^{-2}}$ laser. The \textgamma-photon production from both Bremsstrahlung and Compton scattering is considered, where their contribution is compared by employing both aluminum and gold targets. For gold targets Bremsstrahlung produces more \textgamma-photons than Compton scattering. On the other hand, Compton scattering contribution dominates for aluminum targets. The cone target benefits \textgamma-photon emission due to the magnetic and electrostatic fields generated within the cone, resulting in enhanced hot electron channeling.

The double-layer target concept is examined for \textgamma-photon emission \citep{2018_GuYJ}, during the target interaction with a $3 \kern0.1em {\times} \kern0.1em 10^{23} \kern0.2em \mathrm{W cm^{-2}}$ intensity laser. The front target layer consists from a $22 \kern0.2em \mathrm{\mu m}$ thick hydrogen plasma of $4.5 \kern0.1em {\times} \kern0.1em 10^{21} \kern0.2em \mathrm{cm^{-3}}$ electron density, followed by a $1 \kern0.2em \mathrm{\mu m}$ thick gold foil. While the laser propagates into the hydrogen region it traps electrons, accelerating them to relativistic energies along the laser axis. Part of the laser is reflected by the gold foil and Doppler-shifted, forming attosecond pulses. Furthermore, deformation of the gold surface acts as a concave mirror. The reflected field, when interacting with the hot electrons propagating towards the laser axis, results in collimated \textgamma-photon emission. For the 3D simulation case, $\kappa_\gamma \approx 1.4 \kern0.2em \%$ is obtained.

A structured target \citep{2021_ZhangL} for enhancing $\kappa_\gamma$ consists of four symmetric carbon microwires attached on an aluminum foil, acting as a reflector to the laser pulse. In that study, optimal 3D PIC simulations use microwires of $12 \kern0.2em \mathrm{\mu m}$ length, $1 \kern0.2em \mathrm{\mu m}$ diameter and $3.5 \kern0.2em \mathrm{\mu m}$ spacing. The simulations employ a $50 \kern0.2em \mathrm{PW}$ laser,  at an intensity of $8 \kern0.1em {\times} \kern0.1em 10^{23} \kern0.2em \mathrm{W cm^{-2}}$. As the laser propagates between the wires it expels electrons, that are then accelerated along the laser propagation direction at cut-off energies of $1.5 \kern0.2em \mathrm{GeV}$. By that time, \textgamma-photon emission is low, attributed to the transversely oscillating electron motion. When the laser pulse is reflected by the aluminum foil, $\kappa_\gamma \approx 27 \kern0.2em \%$ is rapidly reached. The simulations reaffirm the appearance of the typical double-lobe structure of the emitted \textgamma-photons.

\begin{figure}[h]
\centering
\includegraphics[width=0.8\linewidth]{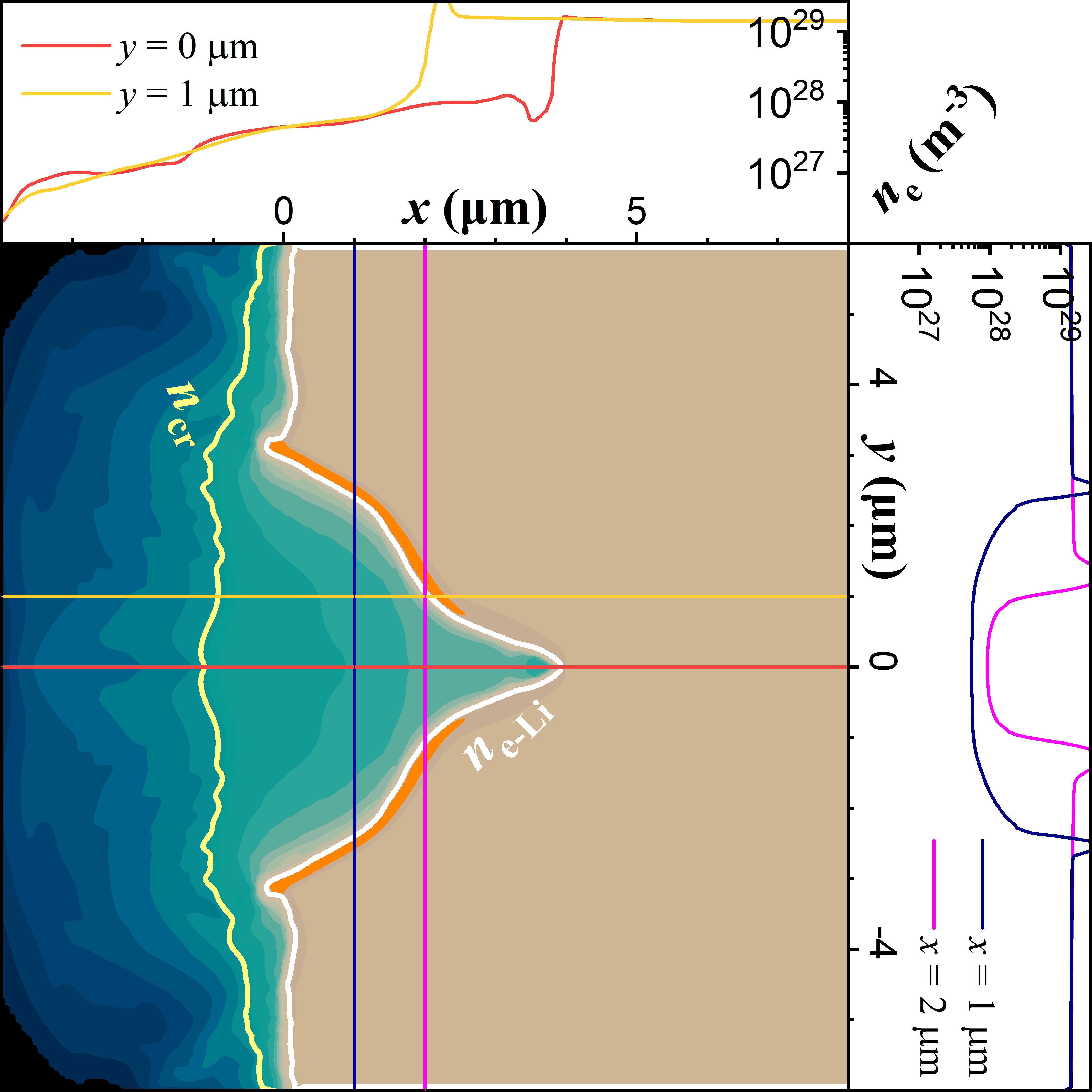}
\caption{Electron number density obtained by magneto-hydrodynamic simulations \citep{2022_TsygvintsevIP} following irradiation of a lithium foil, used as initial conditions for PIC simulations \citep{2022_HadjisolomouP_b}. The yellow contour line is at the critical density and the white contour line is at the lithium solid electron density. The orange saturated contour is overcritical for laser intensities above $10^{23} \kern0.2em \mathrm{W \kern0.1em cm^{-2}}$. Reprinted figure \citep{2022_HadjisolomouP_b} reproduced with permission from P. Hadjisolomou, T. M. Jeong and S. V. Bulanov, Towards bright gamma-ray flash generation from tailored target irradiated by multi-petawatt laser, Sci. Rep., \textbf{12}, 1, 17143, 2022. Copyright 2022 by the Springer Nature Group.}
\label{fig:fig1}
\end{figure}

\par 
A flat foil irradiated by a low amplitude field (several orders of magnitude lower than the main laser pulse) is proposed for enhancing \textgamma-photon emission \citep{2022_HadjisolomouP_b}. The foil irradiation can be attributed either to the laser amplified spontaneous emission or to a secondary low power laser. Magneto-hydrodynamic simulations demonstrate that if for a specific target material the appropriate secondary field is chosen, then a controllable conical-like cavity is tailored on the target \citep{2022_TsygvintsevIP}. In addition to cavitation of the target, an exponential-like preplasma distribution is formed on the target front region. The resulting density distributions, as shown in figure \ref{fig:fig1}, are used as initial conditions in 3D PIC simulations. The simulations are performed at several power levels, at the range of $1 \kern0.2em \mathrm{PW} - 10 \kern0.2em \mathrm{PW}$. The proposed scheme benefits from the combination of isolated aspects previously studied, as near-critical density targets \citep{2016_StarkDJ}, conical targets \citep{2022_ChintalwadS} and preplasma effects \citep{2018_LezhninKV}.

\begin{figure}[h]
\centering
\includegraphics[width=0.8\linewidth]{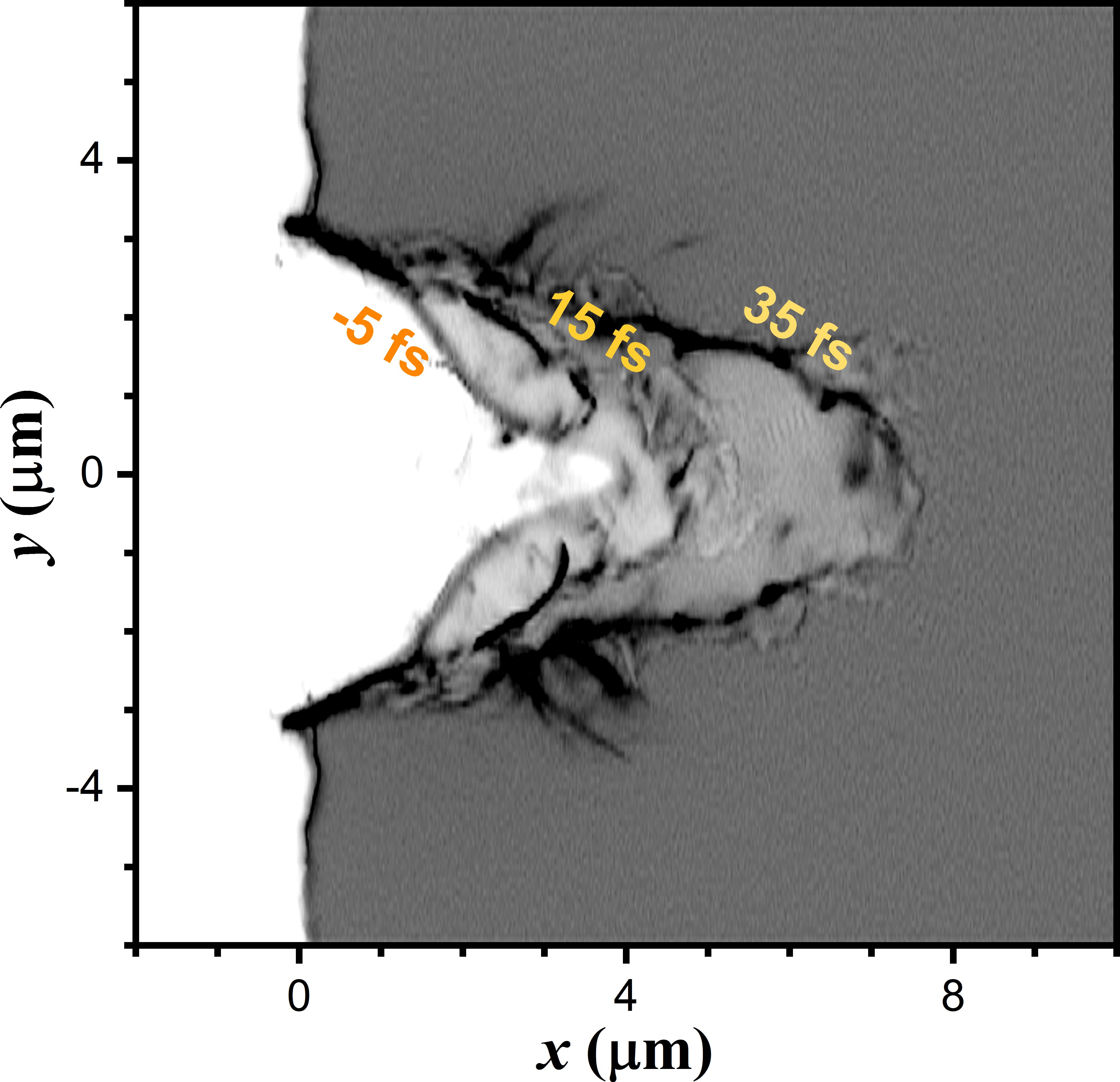}
\caption{Overlay of three successive electron number density distributions, with a time step of $20 \kern0.2em \mathrm{fs}$. The first layer is at $-5 \kern0.2em \mathrm{fs}$, when the main pulse is within the cavity. Reprinted figure \citep{2022_HadjisolomouP_b} reproduced with permission from P. Hadjisolomou, T. M. Jeong and S. V. Bulanov, Towards bright gamma-ray flash generation from tailored target irradiated by multi-petawatt laser, Sci. Rep., \textbf{12}, 1, 17143, 2022. Copyright 2022 by the Springer Nature Group.}
\label{fig:fig2}
\end{figure}

The tailored conical-like cavity is of diameter comparable to the focal spot diameter. As a result, most of the laser field is confined within the cavity, increasing its intensity by an order of magnitude. Although for a $10 \kern0.2em \mathrm{PW}$ laser an intensity of $2.8 \kern0.1em {\times} \kern0.1em  10^{23} \kern0.2em \mathrm{W cm^{-2}}$ is expected on focus, intensities of $2.6 \kern0.1em {\times} \kern0.1em  10^{24} \kern0.2em \mathrm{W cm^{-2}}$ are recorded in the simulations, reaching the regime of prolific \textgamma-photon generation by already existing lasers. Moreover, reflection of the incident laser by the cavity walls results in the appearance of a strong longitudinal electric field component, that assists further cavitation of the foil, as seen in figure \ref{fig:fig2}, allowing existence of ultraintense fields for an extended time. The extreme intensities reached fall in the relativistically transparent regime for lithium. As a result, the laser field penetrates deep into the target, depleting almost all of its energy. For the $10 \kern0.2em \mathrm{PW}$ case, a $\kappa_\gamma \approx 30 \kern0.2em \%$ is reported for a lithium tailored target, decreasing to $\kappa_\gamma \approx 19 \kern0.2em \%$ for a flat foil target. In both cases, the emitted \textgamma-photons are in the GeV-level. Moreover, if a material denser than lithium is used then $\kappa_\gamma$ decreases since the laser-target interaction moves out of the relativistically transparent regime.

\begin{figure}[h]
\centering
\includegraphics[width=0.8\linewidth]{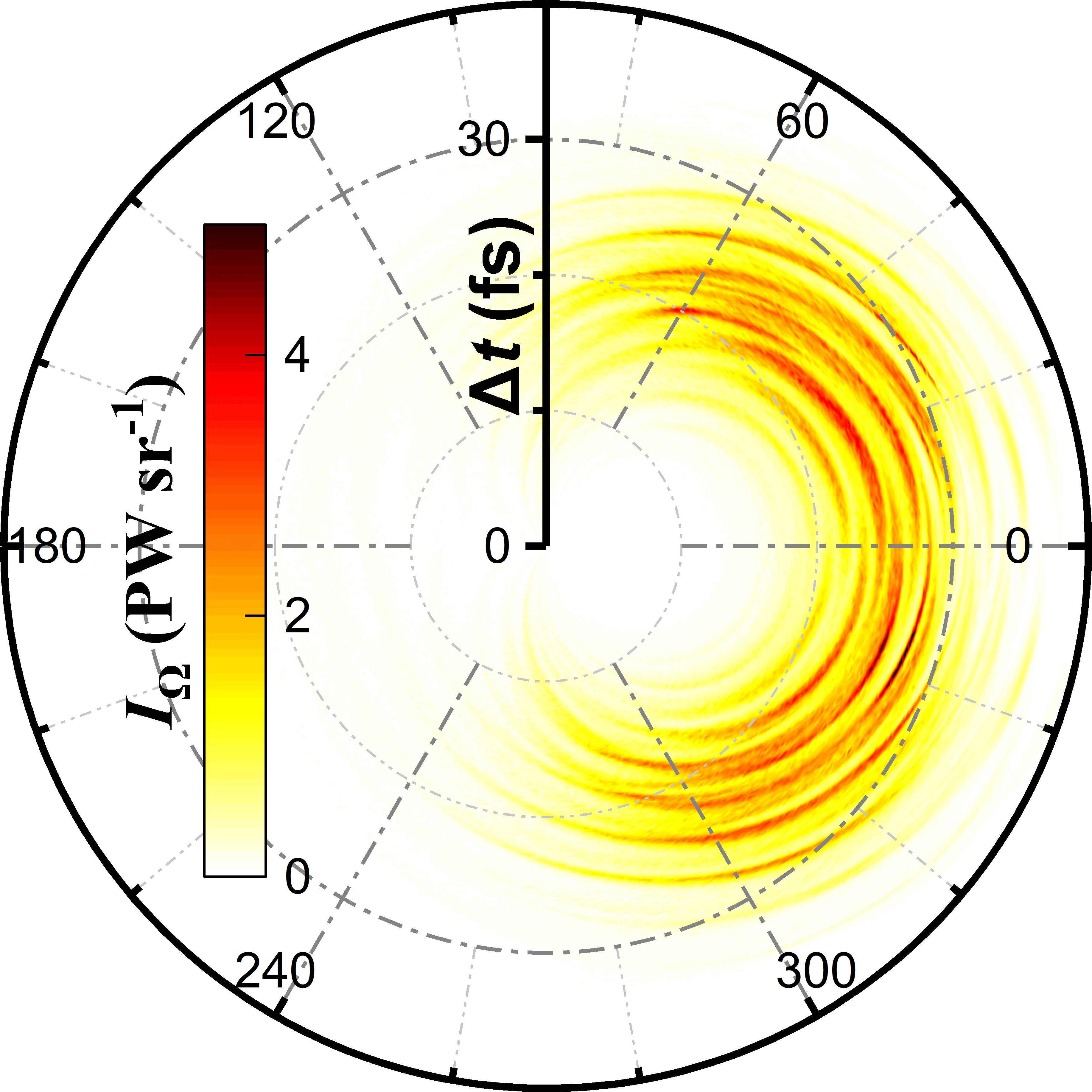}
\caption{Radiant intensity of $\gamma$-photons with respect to the polarization plane, as a function of time-of-flight difference. Reprinted figure \citep{2022_HadjisolomouP_b} reproduced with permission from P. Hadjisolomou, T. M. Jeong and S. V. Bulanov, Towards bright gamma-ray flash generation from tailored target irradiated by multi-petawatt laser, Sci. Rep., \textbf{12}, 1, 17143, 2022. Copyright 2022 by the Springer Nature Group.}
\label{fig:fig7}
\end{figure}

The double-lobe \textgamma-photon emission pattern is verified, with emphasis given on both the spatial and temporal structure of the \textgamma-flash. The \textgamma-photon emission angle is found to be maximized at angles within the range $37^o-55^o$. At all laser power levels the tailored targets (compared to foil targets) give optimal emission angles closer to the laser axis, while the emission angle decreases for increasing laser power. Temporal analysis of the \textgamma-flash reveals a direct connection of \textgamma-photon emission with the peaks of the laser field, as shown in figure \ref{fig:fig7}. Two symmetric emission patterns (with respect to the laser axis) are observed at the two laser polarization hemi-planes, shifted by half wavelength and being suppressed along the laser axis. This pattern is explained by \textgamma-photon emission by electrons co-moving (at a certain angle) with the laser pulse. When the electron momentum angle approaches zero then \textgamma-photon emission is suppressed, forming the double-lobe structure.
\subsection{Multiple Laser Beams Schemes}

For the case of a plane electromagnetic wave, the parameter $\chi_e$ that controls the \textgamma-photon emission (see section \ref{chieg}) is maximized for a charged particle if its momentum is antiparallel to the wave propagation direction, and zero if it is parallel. This observation naturally brought into the discussion the use of multiple colliding lasers to achieve higher $\chi_e$ compared to single beam schemes where in the latter case electrons and positrons are often primarily accelerated in the direction of propagation of the laser.

Interaction of two antiparallel $12.5 \kern0.2em \mathrm{PW}$ power, $4 \kern0.1em {\times} \kern0.1em  10^{23} \kern0.2em \mathrm{cm^{-2}}$ intensity lasers with an aluminum foil is proposed \citep{2015_LuoW}. The dual-beam foil irradiation results in symmetric target compression, reducing the number of hot electrons escaping the target. The dual-beam case corresponds to $\kappa_\gamma = 20.2 \kern0.2em \%$, an increase of a factor of three compared to the  corresponding single-beam case. Overlapping of the incident and reflected pulses results in standing waves, altering $\chi_e$. Moreover, symmetric compression shapes the foil to a concave mirror, allowing higher reflected laser intensities. In addition, electrons from either side of the foil cross over to the opposite side, counter-propagating to the opposite laser. The aforementioned factors enhance \textgamma-photon emission. A similar setup proposes an orthogonal set of polarisations \citep{2015_ZhangP}. The target consists of a $1 \kern0.2em \mathrm{\mu m}$ thick, relativistically overcritical foil. When QED effects are ignored, the two pulses eventually bore through the foil. When QED effects are included, however, the two pulses do not bore through the foil. This is attributed to enhanced laser absorption resulting from radiation reaction.

3D simulations of two antiparallel linearly polarized lasers interacting with a liquid hydrogen slab are compared at the levels of $a_0 = 1000$ and $a_0 = 2000$ \citep{2016_GrismayerT}. This setup results in an enhancement of \textgamma-photon emission, which is attributed to formation of standing waves by the overlapping lasers. For $a_0 = 1000$ the \textgamma-photons are emitted mainly at $\pm 90^o$, while for $a_0 = 2000$, they are emitted at $\pm 30^o$ and $\pm 150^o$.

Combination of the 3D dual-beam scheme with structured targets is also considered \citep{2016_ZhuXL}. Each $3 \kern0.1em {\times} \kern0.1em  10^{22} \kern0.2em \mathrm{W cm^{-2}}$ intensity laser is incident on a conical target filled with nearcritical plasma. The use of this target geometry exhibits a laser intensification of approximately 10 times. The laser propagating in the nearcritical plasma enters the radiation trapping regime \citep{2014_JiLL}. As the damping force equals the Lorentz force then electrons pile up on the laser axis. The result of electron bunching is the induction of a poloidal self-generated magnetic field that keeps electrons on axis. This electron population propagates along with the laser pulse in an oscillatory motion emitting \textgamma-photons. When the two pulses collide with the opposite electron population, further \textgamma-photon emission occurs.

Collision of two counter-propagating circularly polarised pulses of same spatial profile but different temporal profile, propagating in a near-critical plasma, is also investigated \citep{2018_ZhuXL}. The 3D scheme reports $\kappa_\gamma \approx 8 \kern0.2em \%$ at an intensity of $2 \kern0.1em {\times} \kern0.1em 10^{22} \kern0.2em \mathrm{W cm^{-2}}$.

The dual-beam scheme is extended in a quadruple-beam scheme \citep{2016_VranicM}. The optimal polarization configuration (of the linearly polarised lasers) for \textgamma-photon emission corresponds to the counter-propagating laser pulses being of the same polarization but polarized orthogonally to the other two laser pulses. Each laser has a spot size significantly larger than the $0.3 \kern0.2em \mathrm{\mu m}$ wide target used. The parameter $a_0$ ranges from 500 to 2000. For $a_0 > 800$ the \textgamma-photon emission growth rate is lower than predicted \citep{2016_GrismayerT}, indicating disruption of the standing waves \cite{2018_LuoW,2019_SladeLowtherC}. For $10^{24} \kern0.2em \mathrm{cm^{-2}}$ intensity, $\kappa_\gamma \approx 50 \kern0.2em \%$, which is attributed to the high \textgamma-photon emission growth rate of the configuration. Most of the \textgamma-photon energy is radiated in a cross-like pattern. This laser configuration has been also considered for an order of magnitude lower intensities and a significantly larger target \citep{2017_GongZ}, where for a $a_0 = 325$ laser a $\kappa_\gamma \approx 53.7 \kern0.2em \%$ is reported.

Interaction of two 3D elliptically polarised lasers with two $0.32 \kern0.2em \mathrm{\mu m}$ thick diamond-like foils is proposed \citep{2017_LiHZ}. Although circularly polarized beams benefit the radiation pressure acceleration mechanism, elliptically polarized lasers reduce the transverse instabilities. The two ellipse axis correspond to $a_0$ values of 237 and 154. The laser-foil interaction forms a conical-like structure in the foil before the pulse peak amplitude interacts with the target. During that stage, an electron population is dragged from the cavity walls and then interacting with the reflected waves, but with a low \textgamma-photon yield due to Doppler shifting of the reflected laser. Following the foil cavitation, the target deforms, lowering its density along the laser axis. If the spacing of the foils is suitably chosen, the time the two pulses penetrate the foils coincides with the time the two foil front surfaces collide. Hot electrons cross over to the opposite pulse region, resulting in high $\chi_e$ values, enhancing \textgamma-photon emission by a factor of four compared to the single-foil interaction under the same conditions.

An alternative to double-foils is the convex target \citep{2020_YasenN}, interacting with a $4 \kern0.1em {\times} \kern0.1em  10^{23} \kern0.2em \mathrm{cm^{-2}}$ intensity laser. The targets consist of $2 \kern0.2em \mathrm{\mu m}$ thick aluminum. The two foils have a separation distance of $2 \kern0.2em \mathrm{\mu m}$, the same as that of the convex targets in the centre. The results indicate an $11 \kern0.2em \%$ \textgamma-photon emission enhancement for the convex targets case, reaching $\kappa_\gamma \approx 13.2 \kern0.2em \%$.

\subsection{The $\lambda^3$ Scheme - An Alternative to Multi-Beam Configurations} \label{import}

\begin{figure}
  \centering
  \includegraphics[width=0.8\linewidth]{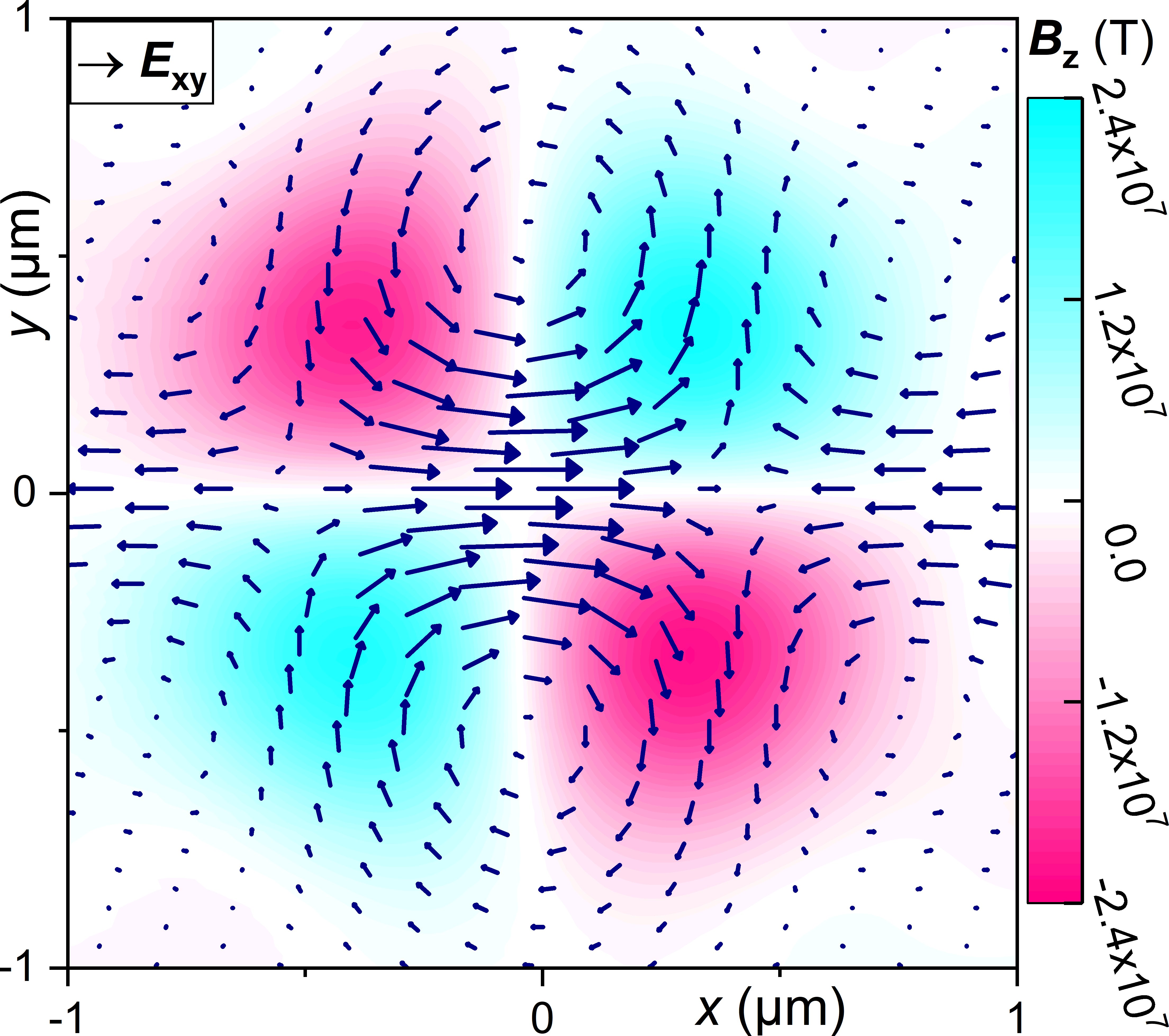}
  \caption{The $\lambda^3$ radially polarized laser fields. The black arrows correspond to the electric field vectors, over-plotted on a contour of the magnetic field. Reprinted figure \citep{2022_HadjisolomouP} reproduced with permission from P. Hadjisolomou, T. M. Jeong, P. Valenta, D. Kolenaty, R. Versaci, V. Olšovcová, C. P. Ridgers, and S. V. Bulanov, Gamma-ray flash in the interaction of a tightly focused single-cycle ultra-intense laser pulse with a solid target, J. Plasma Phys., \textbf{88}, 1, 905880104, 2022. Copyright 2022 by Cambridge University Press.}
\label{fig:Fields}
\end{figure}

The usage of multi-beam configurations significantly enhances \textgamma-photon emission from the interaction of intense lasers with matter. The $4\pi$-spherical focusing scheme can be regarded as an extreme case of either multiple laser beam focusing or tight focusing scheme \citep{2020_JeongTM}. Focusing a large number of single-cycle pulses to a single spot mimics the $\lambda^3$ regime \citep{2002_MourouG}, that provides the highest possible laser intensity for the least energy, at a given power. The utility of the $\lambda^3$ regime for studying QED effects during the interaction of an ultraintense laser with a target has recently been demonstrated \citep{2021_HadjisolomouP, 2022_HadjisolomouP} in a 3D multiparametric study, where a sample case of the $\lambda^3$ laser fields used are shown in figure \ref{fig:Fields}. The varying parameters include the laser polarization, target thickness, target electron density and laser power.

\begin{figure}
  \centering
  \includegraphics[width=1.0\linewidth]{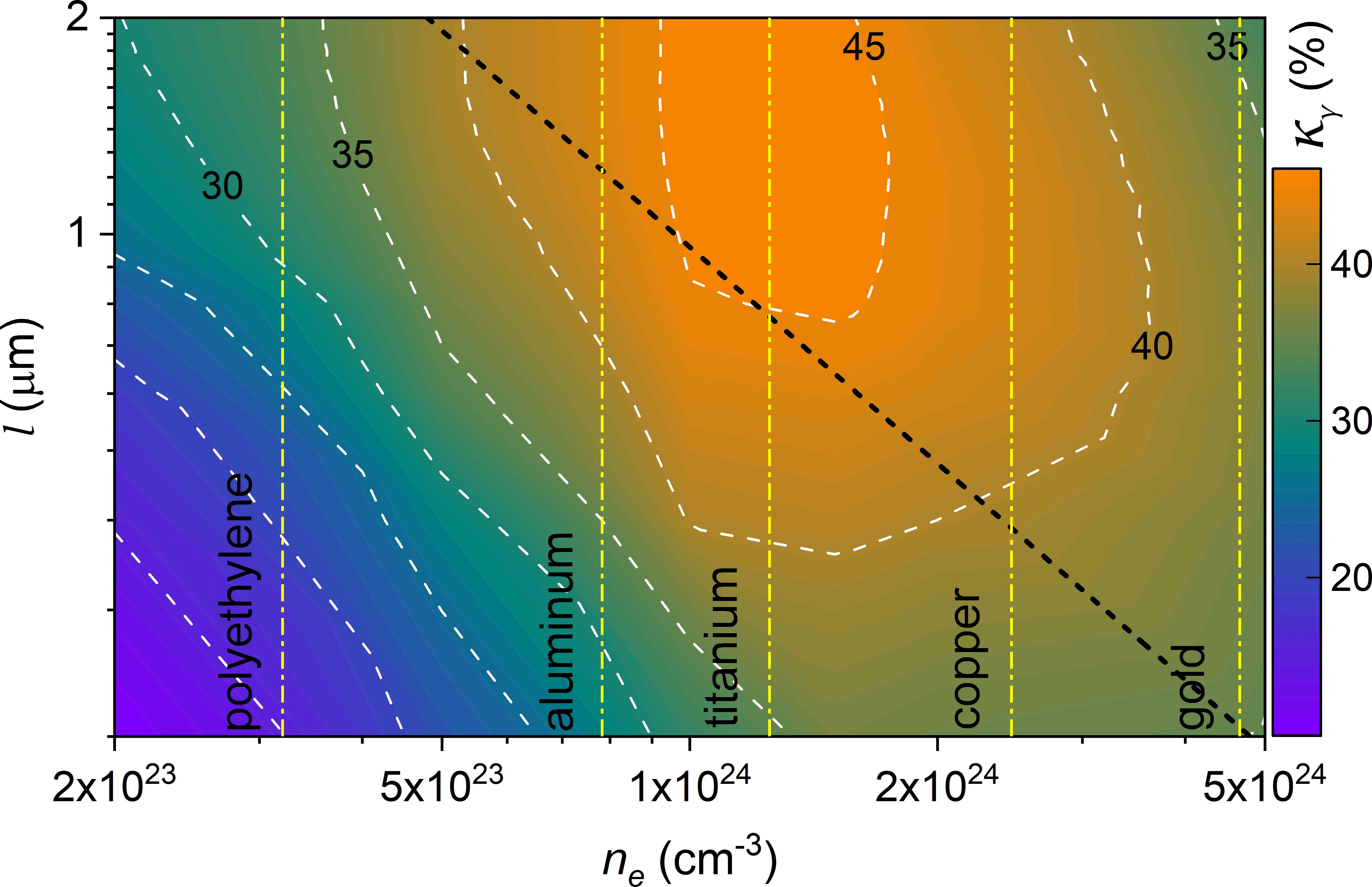}
  \caption{$\kappa_\gamma$ as a function of the target thickness and electron density \citep{2021_HadjisolomouP}. The black dashed line follows the relation between target thickness and electron density for optimal laser coupling to the target \citep{1998_VshivkovVA}. Reprinted figure \citep{2021_HadjisolomouP} reproduced with permission from P. Hadjisolomou, T. M. Jeong, P. Valenta, G. Korn and S. V. Bulanov, Gamma-ray flash generation in irradiating a thin foil target by a single-cycle tightly focused extreme power laser pulse, Phys. Rev. E, \textbf{104}, 1, 015203, 2021. Copyright 2021 by the American Physical Society.}
  \label{fig:3}
\end{figure}

The simulations suggest substantial enhancement in the emission of \textgamma-photons, compared to multi-beam ($\geq 4$) schemes. At a laser intensity of $10^{25} \kern0.2em \mathrm{cm^{-2}}$, corresponding to $a_0 \approx 2700$ and reached by an $80 \kern0.2em \mathrm{PW}$ laser, optimal conditions for \textgamma-photon emission occur for a radially polarized laser combined with a $2 \kern0.2em \mathrm{\mu m}$ thick titanium target. Under these conditions, a $\kappa_\gamma \approx 47 \kern0.2em \%$ is recorded, as shown in figure \ref{fig:3}. After parameter optimization, a linearly polarized laser gives $\kappa_\gamma \approx 42 \kern0.2em \%$ and an azimuthally polarized laser (magnetic field rotates around the laser propagation axis) gives $\kappa_\gamma \approx 29 \kern0.2em \%$.

The high $\kappa_\gamma$ for radially polarized lasers is attributed to the dominant longitudinal field component that increases the laser coupling to the target \citep{1998_VshivkovVA}. Although the intensity of a linearly polarized laser when tightly focused is double that of a radially polarized laser, the former has a smaller longitudinal electric field component \citep{2018_JeongTM}. By definition, for the azimuthally polarized laser no longitudinal electric field exists. This field component drives the formation of a cavity in the foil. The cavity is symmetric and has a conical-like form for the radially polarized laser case. The case of linearly polarized laser corresponds to a square-like cavity. The cavity formation is significantly delayed for the azimuthally polarized laser case. In the case of the radially polarized laser case, reflection, diffraction and interference of the laser field in the cavity results in a peak intensity of $8.8 \kern0.1em {\times} \kern0.1em 10^{25} \kern0.2em \mathrm{W cm^{-2}}$.

\begin{figure}
  \centering
  \includegraphics[width=1.00\linewidth]{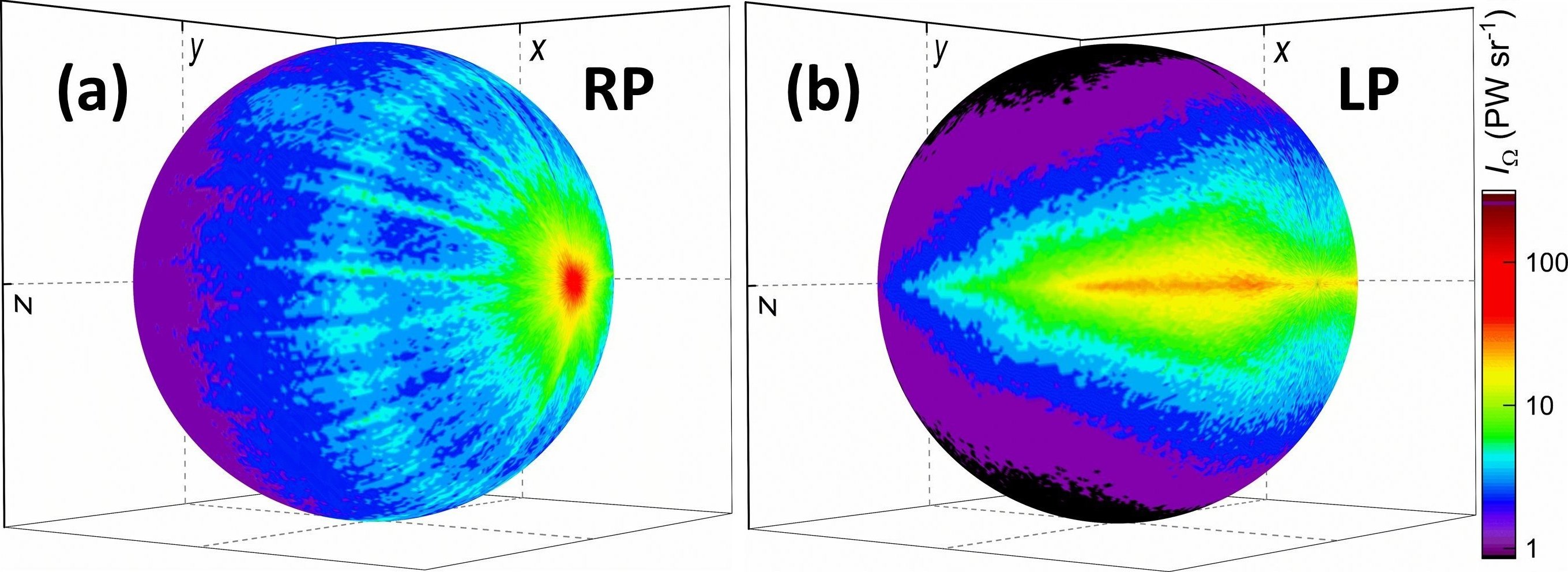}
  \caption{\textgamma-photon radiant intensity for (\textit{a}) a RP laser and (\textit{b}) a LP laser laser. Reprinted figure \citep{2022_HadjisolomouP} reproduced with permission from P. Hadjisolomou, T. M. Jeong, P. Valenta, D. Kolenaty, R. Versaci, V. Olšovcová, C. P. Ridgers, and S. V. Bulanov, Gamma-ray flash in the interaction of a tightly focused single-cycle ultra-intense laser pulse with a solid target, J. Plasma Phys., \textbf{88}, 1, 905880104, 2022. Copyright 2022 by Cambridge University Press.}
\label{fig:Flash}
\end{figure}

For the radially polarized laser case, several distinct energetic electron populations are identified. These groups are directed at $0^o$, $180^0$ and $60^o$ and are characterized by the quarter-period. The emission of high energy \textgamma-photons is identified along the same directions, during the same time instances. The radially polarised laser case exhibits an extremely collimated \textgamma-photon population along the laser axis, a secondary population on the opposite direction and a third but minor population at $60^0$, as seen in figure \ref{fig:Flash}. For the linearly polarized laser case, a double-lobe structure is observed. For the azimuthally polarized laser case, the \textgamma-photon emission is more isotropic.

The \textgamma-photons propagate ballistically within an expanding spherical shell. By calculating the \textgamma-photon standard deviation, it is estimated that the radially, linearly and azimuthally polarized lasers correspond to \textgamma-flashes of $31 \kern0.2em \mathrm{PW}$, $28 \kern0.2em \mathrm{PW}$ and $13 \kern0.2em \mathrm{PW}$ respectively. Variation on the laser power reveals that $\kappa_\gamma$ follows a saturating function. By decreasing the target density the threshold for observing \textgamma-photon emission shifts to lower power values.
\section{Radioactive nuclei and electron-positron pair production by irradiating high-Z target with $\gamma$-photons}

FLUKA \citep{2022_AhdidaC, 2015_BattistoniG} Monte-Carlo simulations are suitable to examine the effect of \textgamma-photons, electrons, positrons and ions interacting with a High-Z target \citep{2022_KolenatyD}. The Monte-Carlo simulations import as primary particles the PIC output (position, momentum and weight) macroparticles [\textgamma-photons, electrons, positrons, titanium ions ($Ti^{+}$)] corresponding to the optimal $\kappa_\gamma$ case \citep{2021_HadjisolomouP} described in section \ref{import}. The number of primary \textgamma-photons approximately equals that of electrons, with $Ti^{+}$ and positrons being one and two orders of magnitude less, respectively. All PIC macroparticles apart from positrons are collimated along the laser axis due to the radially polarized laser used.

A $100 \kern0.2em \mathrm{mm}$ diameter lead (${}^{207.19}Pb$) cylindrical target is placed $1 \kern0.2em \mathrm{mm}$ away from the focal spot. Lead, due to its high Z-number, benefits from high cross-section for $e^-e^+$ pair generation and for the giant dipole resonance (GDR). The target thickness varies from $1 \kern0.2em \mathrm{mm}$ to $100 \kern0.2em \mathrm{mm}$. The energy deposited by the primary particles saturates for a $100 \kern0.2em \mathrm{mm}$ thick target.

When \textgamma-photons collide with the target the ratio of the number of the emitted electrons to positrons is approximately unity, attributed to $e^-e^+$ pair generation. Positron generation is optimal for a $5 \kern0.2em \mathrm{mm}$ thick target, with the number of available positrons increased by a factor of 16 compared to the primary positrons. Target irradiation by the directional primary particles is reflected in the generation of collimated positrons.

Neutrons are produced via photonuclear reactions. The interactions are either direct (by initial \textgamma-photons) or indirect (by Bremsstrahlung secondary \textgamma-photons). For lead, photonuclear interactions cross-section peaks at $13 \kern0.2em \mathrm{MeV}$. High neutron fluxes are required for neutron diffraction \citep{2020_VogelSC}, neutron resonance spectroscopy \citep{2010_HigginsonDP, 2020_ZimmerM} and nuclear waste management \citep{2001_RevolJP}.

\begin{figure}
  \centering
  \includegraphics [width=1.0\linewidth]{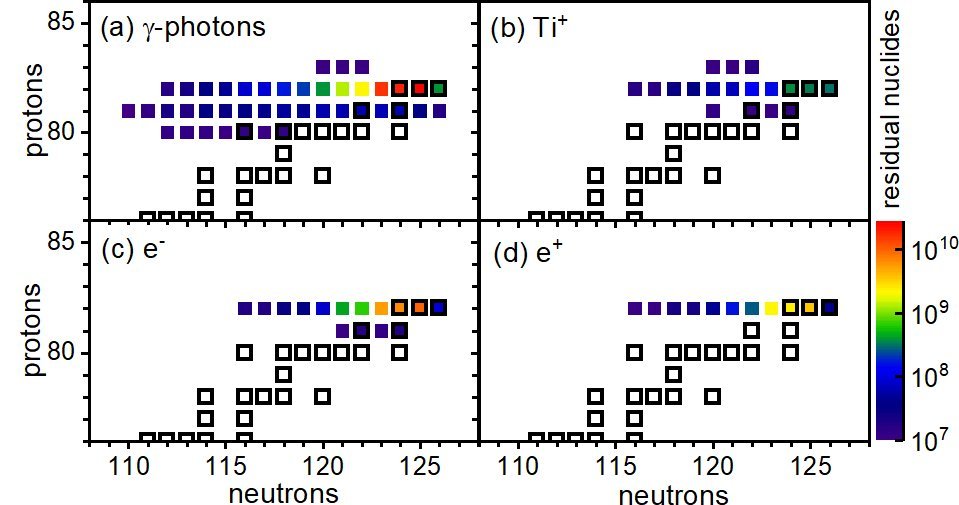}
  \caption{Chart of residual nuclides obtained from MC simulation per $\lambda^3$-pulse \citep{2022_HadjisolomouP, 2022_KolenatyD}. Independent contribution from each primary particle specie is presented. Stable nuclides are highlighted with a black frame. Reprinted figure \citep{2022_HadjisolomouP} reproduced with permission from P. Hadjisolomou, T. M. Jeong, P. Valenta, D. Kolenaty, R. Versaci, V. Olšovcová, C. P. Ridgers, and S. V. Bulanov, Gamma-ray flash in the interaction of a tightly focused single-cycle ultra-intense laser pulse with a solid target, J. Plasma Phys., \textbf{88}, 1, 905880104, 2022. Copyright 2022 by Cambridge University Press.}
  \label{fig:FLUKAactivation}
\end{figure}

In the $100 \kern0.2em \mathrm{mm}$ thick target, residual nuclides are produced via photonuclear reactions, as seen in figure \ref{fig:FLUKAactivation}. The produced ${}^{203}Pb$ decays via electron capture to ${}^{203}Tl$, with a half-life of $52 \kern0.2em \mathrm{h}$. Since no hadron is involved in the decay, ${}^{203}Pb$ suits medical imaging \citep{2014_AzzamA}. Similarly, the produced ${}^{201}Tl$ decays via electron capture to ${}^{201}Hg$, also suitable for medical imaging \citep{1999_TadamuraE}.

\par 
\section{All-optical nonlinear Breit-Wheeler pair production}

An important future application of high-energy \textgamma-photons generated with multi-PW laser systems is the study of photon-photon interactions in quantum electrodynamics (see the recent review \citet{2023_FedotovA}). In particular, nonlinear Breit-Wheeler pair production \citep{1934_BreitG, 1962_ReissH, 1964_NikishovA, 1966_YakovlevVP} requires large numbers of high-energy \textgamma-photons to overcome the pair creation energy threshold and provide a sufficient signal to enable experimental observation.

\begin{figure}
  \centering
  \includegraphics [width=1.0\linewidth]{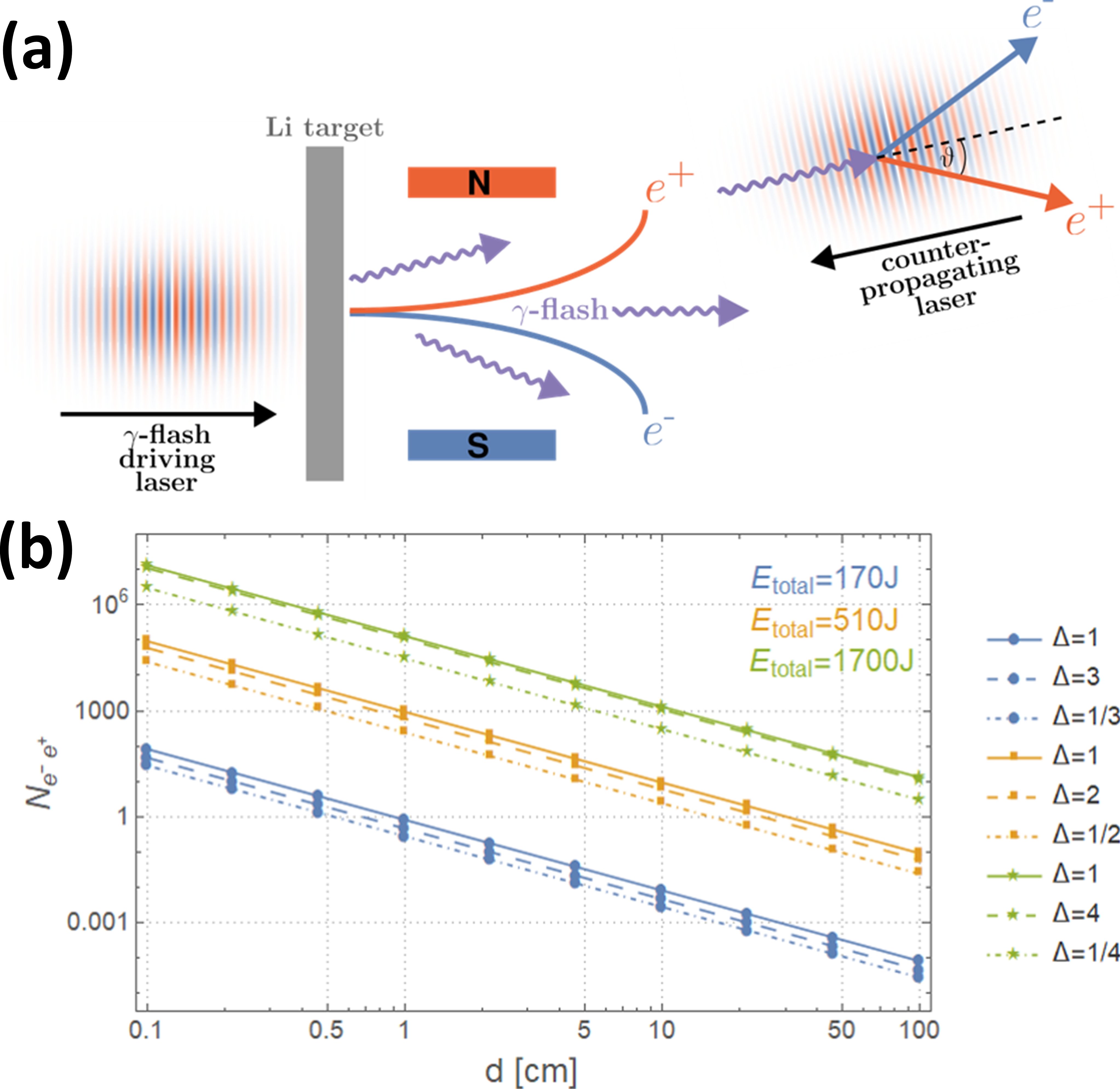}
  \caption{(\textit{a}) The proposed all-optical nonlinear Breit-Wheeler pair production setup. (\textit{b}) The total number of $e^-e^+$ pairs expected per shot for each $E_{\text{total}}$ as a function of the the distance, $d$, from the photon point source at the lithium target rear surface to the focus of the secondary laser pulse. Reprinted figure \citep{2023_MacLeodAJ} reproduced with permission from A. J. MacLeod, P. Hadjisolomou, T. M. Jeong S. V. and Bulanov, All-optical nonlinear Breit-Wheeler pair production with $\ensuremath{\gamma}$-flash photons, Phys. Rev. A, \textbf{107}, 1, 012215, 2023. Copyright 2023 by the American Physical Society.}
  \label{fig:AllOptical}
\end{figure}

The increase in both the conversion efficiency and maximum photon energy of the \textgamma-ray flash with structured targets was recently exploited in \citet{2023_MacLeodAJ} to provide a simple all-optical setup for experimentally studying nonlinear Breit-Wheeler pair production at current and next generation high-power laser facilities. The proposed setup consisted of two stages, shown schematically in figure \ref{fig:AllOptical}(a). Firstly, a lithium target is irradiated with a combination of a nanosecond-duration prepulse and a high-intensity main laser pulse. The nanosecond-duration prepulse generates a conical channel in the lithium target which leads to an intensity enhancement of the main pulse, subsequently resulting in increase of the total yield and energy of \textgamma-photons produced, as described above in section \ref{ST}. Secondly, the produced \textgamma-photons collide with a secondary counter-propagating high-intensity laser pulse to produce $e^-e^+$ pairs via the nonlinear Breit-Wheeler process.

Synchronization between the two stages can be achieved by splitting the total available laser pulse energy $E_{\text{total}}$ into two beams, one to drive the \textgamma-photon production with pulse energy $E_{\text{flash}}$ and one to generate $e^-e^+$ pairs via the nonlinear Breit-Wheeler process with pulse energy $E_{\text{pairs}}$. Three different cases for the total available laser energy were considered, $E_{\text{total}} = (170, 510, 1700) \kern0.2em \mathrm{J}$, which for a pulse duration of $17 \kern0.2em \mathrm{fs}$ corresponds to a power of $(10, 30, 100) \mathrm{PW}$, respectively.

Figure \ref{fig:AllOptical}(b) shows the total number of $e^-e^+$ pairs expected per shot for each $E_{\text{total}}$ as a function of the the distance, $d$, from the photon point source at the lithium target rear surface to the focus of the secondary laser pulse. In each case, the ratio $\Delta = E_{\text{flash}}/E_{\text{pairs}}$, gives the relative splitting of the total available pulse energy into the two stages of the setup, where the optimal number of $e^-e^+$ pairs occurs for the case $\Delta = 1$. The total number of $e^-e^+$ pairs with a propagation distance $d = 10 \kern0.2em \mathrm{cm}$ was shown to offer comparable numbers of pairs per shot to alternative schemes based on Bremsstrahlung photon generation (see e.g. \cite{2022_GolubA}).


\section{Summary and Conclusions}

The discussion of QED effcts in laser-matter interactions occupies a large volume of the laser plasma physics literature \citep{2006_MourouGA, 2009_EhlotzkyF, 2012_DiPiazzaA, 2015_BulanovSV, 2020_ZhangP, 2021_FedotovA, 2022_GonoskovA}. The proportion of the literature devoted on computational description of those effects increases alongside available computational power. In this article we present recent results of the QED PIC literature.

We distinguish interactions of ultra-intense lasers with targets, both in the context of targets with preplasma and structured targets, both exhibiting a prolific \textgamma-photon emission. Often, the suggested targets increase \textgamma-photon emission by increasing the laser intensity, as for example by near-critical, conical, or tailored targets.

Emission of \textgamma-photons is suppressed if the emitting electron moves in the same direction as the laser. As a result, multi-beam setups are suggested, demonstrating significant enhancement of \textgamma-photon emission. An alternative to the multi-beam setup is the tight focusing scheme. This scheme, generalized in the $\lambda^3$ regime, demonstrates $\kappa_\gamma$ values only comparable to to that of four-beam (or higher) configurations.

The results of the $\lambda^3$ QED PIC simulations combined with Monte-Carlo simulations illustrate a drastic effect of the primary laser-target interaction on high-Z materials, with significant increase of the generated positron number by the Bethe-Heitler process and strong activation of the high-Z material mainly due to photonuclear reactions. Moreover, one can harness the \textgamma-ray flash properties to demonstrate the Breit-Wheeler pair production process based in an all-optical setup.


\begin{acknowledgments}
The authors would like to acknowledge useful communication with A. S. Pirozhkov, T. Z. Esirkepov, M. Kando and T. Kawachi. This work is supported by the project “Advanced research using high intensity laser produced photons and particles” (ADONIS) \allowbreak{(CZ.02.1.01/\allowbreak0.0/0.0/16 019/0000789)} from the European Regional Development Fund. C. P. Ridgers would like to acknowledge support from the UK Engineering and Physical Sciences Research Council, grant number EP/V049461/1.
\end{acknowledgments}

\section*{Data Availability Statement}
Data sharing is not applicable to this article as no new data were created or analyzed in this study.

\bibliography{biblio}

\end{document}